\documentclass{emulateapj}

\usepackage{amssymb, amsmath, apjfonts, natbib, color}

\def\degree{{\circ}}
\newdimen\digitwidth
\setbox1=\hbox{0}
\digitwidth=\wd1
\catcode`"=\active
\def"{\kern\digitwidth}

\slugcomment{Accepted for Publication in ApJL}

\begin{document}

\title{THE VERTICAL X-SHAPED STRUCTURE IN THE MILKY WAY: \\EVIDENCE FROM A SIMPLE BOXY BULGE MODEL}

\author{Zhao-Yu Li\altaffilmark{1} and Juntai Shen\altaffilmark{1, 2}}

\altaffiltext{1}{Key Laboratory for Research in Galaxies and Cosmology, Shanghai Astronomical Observatory, Chinese Academy of Sciences, 80 Nandan Road, Shanghai 200030, China}
\altaffiltext{2}{Correspondence should be addressed to Juntai Shen: jshen@shao.ac.cn}
\begin{abstract}
A vertical X-shaped structure in the Galactic bulge was recently reported. 
Here we present evidence of a similar X-shaped structure in the Shen et
al. (2010) bar/boxy bulge model that simultaneously matches the stellar
kinematics successfully. The X-shaped structure is found in the central region
of our bar/boxy bulge model, and is qualitatively consistent with the observed
one in many aspects. End-to-end separations of the X-shaped structure
in the radial and vertical directions are roughly 3 kpc and 1.8 kpc,
respectively. The X-shaped structure contains about 7\% of light in the boxy bulge region, but it is significant enough to be identified in observations.
An X-shaped structure naturally arises in the formation of bar/boxy bulges, and is mainly associated with orbits trapped around the vertically-extended $x_1$ family. Like the bar in our model, the X-shaped structure tilts away from the Sun--Galactic center line by $20^\degree$. The X-shaped structure becomes increasingly symmetric about the disk plane, so the observed symmetry may indicate that it formed at least a few billion years ago. The existence of the vertical X-shaped structure suggests that the formation of the Milky
Way bulge is shaped mainly by internal disk dynamical instabilities.

\end{abstract}


\section{INTRODUCTION}

Understanding the Galactic structure is non-trivial, mostly because we are
located in the disk plane. Infrared imagery shows that the Milky Way contains a boxy, parallelogram-shaped bulge  (Maihara et al. 1978; Weiland et al. 1994). This can be explained by a tilted bar; the near end of the bar is closer to us than the far side, consequently it appears to be bigger than the other side (Blitz \& Spergel 1991). A good distance indicator for structures of the Galaxy is red clump (RC) stars because their luminosity depends weakly on the stellar mass, age and metallicity (Stanek \& Garnavich 1998). Studies of the asymmetric distribution of RC in the bulge region suggested that the bar  probably extends $\sim 20^\degree - 30^\degree$ from the Sun--Galactic center (GC) line (Stanek et al. 1994, 1997). The detailed properties of the Galactic bar are still under active debate (e.g., Sevenster et al. 1999; Beaulieu et al. 2000; Bissantz \& Gerhard 2002; Bissantz et al. 2003; Bissantz et al. 2004; Babusiaux \& Gilmore 2005; Benjamin et al. 2005; Cabrera-Lavers et al. 2007; Rattenbury et al. 2007; Martinez-Valpuesta \& Gerhard 2011; Gerhard \& Martinez-Valpuesta 2012).

Recently, two groups independently reported the bimodal brightness
distribution of the RC in the Galactic bulge (McWilliam \& Zoccali 2010 hereafter MZ10; Nataf et al. 2010). MZ10 suggested that the bimodality is hard to explain with a tilted bar since the line of sight crossing the bar can only result in stars with one distance. One possibility speculated by Nataf et al. (2010) is that one RC population belongs to the bar and the other to the spheroidal component of the bulge. 
Another puzzling fact is that distances of the bright and faint RC are roughly constant at different latitudes, which was hard to understand with a naive straight bar. They proposed that these observed evidences can be well explained with a vertical X-shaped structure in the bulge region. The existence of this particular structure is later verified by Saito et al. (2011) (hereafter S11). They found that the X-shaped structure exists within (at least) $|l| \leq 2^\degree$, and has front-back symmetry. 

Observationally, about half of edge-on disk galaxies have boxy/peanut-shaped
(BPS) bulges (L\"utticke et al. 2000). The high fraction of BPS bulges is very similar to the bar fraction (Eskridge et al. 2000; Men\'endez-Delmestre et al. 2007; Marinova \& Jogee 2007; Aguerri et al. 2009), hinting for a possible connection between BPS bulges and bars (Bureau \& Freeman 1999; Merrifield \& Kuijken 1999; Laurikainen et al. 2011). Numerical simulations have long found that evolved bars usually appear to be boxy/peanut-shaped when viewed side-on (e.g., Combes \& Sanders 1981; Raha et al. 1991; Athanassoula 2005; Martinez-Valpuesta, Shlosman \& Heller 2006). On the other hand, complicated structures are frequently found in extragalactic BPS bulges, such as centered or off-centered X structures (Bureau et al. 2006). However, since significant image processing is usually required to highlight the faint extragalactic X-shaped structures, MZ10 was unsure whether or not they are the convincing counterpart of the Galactic X-shaped structure.

The Bulge Radial Velocity Assay (BRAVA) uses M giants to probe stellar kinematics of the Galactic bulge (Rich et al. 2007; Howard et al. 2008; Kunder et al. 2012). In addition to photometric studies, stellar kinematics can provide important dynamical constraints to better understand the Galactic bulge. Howard et al. (2009) found a strong cylindrical rotation in the Galactic bulge, which is hard to explain with a classical spheroidal component. Shen et al. (2010) further constructed a simple but realistic Milky Way boxy bulge model, where a dynamically cold disk self-consistently develops a bar. The bar quickly buckles and thickens in the vertical direction due to the buckling/firehose instability (Toomre 1966; Raha et al. 1991). As seen from Sun, the thickened part of the bar appears as the boxy bulge of our Galaxy. More importantly, the model matches detailed stellar kinematics of BRAVA strikingly well with no need for a significant classical bulge component. 

The motivation of this work is to test whether or not an X-shaped structure exists in the Shen et al. (2010) bar/boxy bulge model, and whether it is significant enough to explain the observed features in the Galactic bulge. As we show in this Letter, the model in Shen et al. (2010) naturally produces a vertical X-shaped structure within the bar. Furthermore, the structure in our model is significant enough to be reliably detected, and its properties are broadly consistent with observations in many aspects. 

\section{THE MILKY WAY BULGE MODEL}

The $N$-body model of Milky Way boxy bulge employed here is identical to that in Shen
et al. (2010). Briefly, the simulation starts 
with one million particles in a thin disk with an exponential surface density
distribution. The initial disk is dynamically cold with Toomre's $Q\sim
1.2$. In this simulation, a bar forms from the disk spontaneously and quickly
buckles in the vertical direction. The structures of this simulated disk
galaxy become roughly steady in the face-on view 
after $\sim$ 2.4 Gyr. The snapshot of this
simulation at 4.8 Gyr, which was also used in Shen et al. (2010) to match the 
stellar kinematics of BRAVA, is selected here to study the disk structures. The
length unit of the simulation is $R_{\rm d, 0}$ = 1.9 kpc, which is the
scale length of the initial exponential disk. We refer the interested reader
to Shen et al. (2010) for more details of the model.

We create a mock image from this snapshot by projecting the particles
from the 3-D space onto a 2-D plane. The pixel
value represents the number of particles projected into the pixel. Such a
mock image has its unique advantages. First, there are no instrumental 
uncertainties, such as read noises, bias subtraction or flat-fielding.
Second, variations of the Point Spread Function (PSF) from atmospheric 
turbulence or from focus changes across the focal plane are absent.
Third, there is no Galactic dust extinction,  foreground star or
background galaxy that may contaminate the light of the main galaxy. In
addition, there is no need for sky subtraction. Perhaps most importantly, we
can project our model in arbitrary viewing angles, enabling a
thorough study on the structures of this disk galaxy.

\section{RESULTS AND DISCUSSIONS}

\subsection{Basic properties of the X-shaped structure}

The edge-on galaxy with a side-on bar is shown in the upper
panel of Figure~1, and an X-shaped structure is discernible in the 
inner region of the boxy bulge. 
The bar length, defined as the distance between the two end points of the bar (2 $R_{\rm bar}$), is about 8 kpc (Shen et al. 2010).
After applying a mask to cover the X-shaped structure, we
use the IRAF task ELLIPSE to fit the elliptical isophote of this
edge-on image. The center is fixed for each isophote, whereas the ellipticity
and position angle are free parameters. Then we use the task BMODEL to construct a
model based on the extracted elliptical isophotes to properly account for the
underlying smooth component of the edge-on galaxy. This model is subtracted
from the original image to produce a residual image (lower-panel of Figure~1),
which more clearly highlights the X-shaped structure. We carried out several tests and
found that this residual image is insensitive to the size of the mask
used. 
The four arms of the X-shaped structure are cone-like (with finite thickness). They are narrow towards the Galactic center, and become wider outward. In the X direction, the end-to-end separation between the inner two edges of the X-shaped structure is $\sim$2 kpc. For the outer two edges, the end-to-end separation is $\sim$4 kpc. We estimate the size of the X-shaped structure in the X direction by averaging the two separations, which yields about $\sim$3 kpc. This value is less than half of the full length of the bar (8 kpc). Similarly, in the Z axis, the end-to-end separation between the inner two edges of the X-shaped structure is $\sim$1.2 kpc. For the outer two edges in the Z axis, this separation is $\sim$2.4 kpc. Therefore the size of the X-shaped structure in the Z direction is $\sim$1.8 kpc. Since the boundaries of this structure are not clear-cut, our measurement has an uncertainty of about 0.2 kpc.
From the image it is also apparent that the X shape is quite
symmetric in the X-Z plane. By summing up the pixels with positive values in the X-shaped region, we estimate that the light fraction of this X-shaped structure relative to the whole boxy bulge region is about 7\%.

\figurenum{1}
\begin{figure}
\epsscale{1.25}
\centerline{\plotone{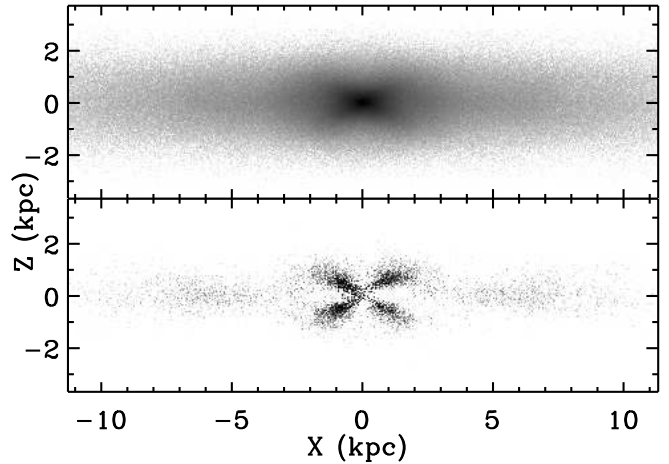}}
\caption{The upper panel shows the side-on view of the bar in our model.  The
  lower panel shows the residual after subtracting the underlying smooth light
  contribution. The vertical X-shaped structure is highlighted in this
  residual image.} 
\end{figure}

To further confirm the existence of this X-shaped structure, we smooth the image with
a median filter and subtract it from the original image. The residual clearly
shows an X-shaped structure in the disk, which is
less extended than that in the lower panel of Figure~1. Because the median
filter technique is only sensitive to small scale and strong structures, the residual image does not reliably reflect the size of the X-shaped
structure. Near the end points of the X-shaped structure the intensity contrast relative to the local background is low, and the structure becomes quite broad. Therefore, it is not very surprising that the extended faint parts of the X-shaped structure do not stand out in the median filter processed residual image.

\figurenum{2}
\begin{figure*}
\epsscale{0.9}
\plotone{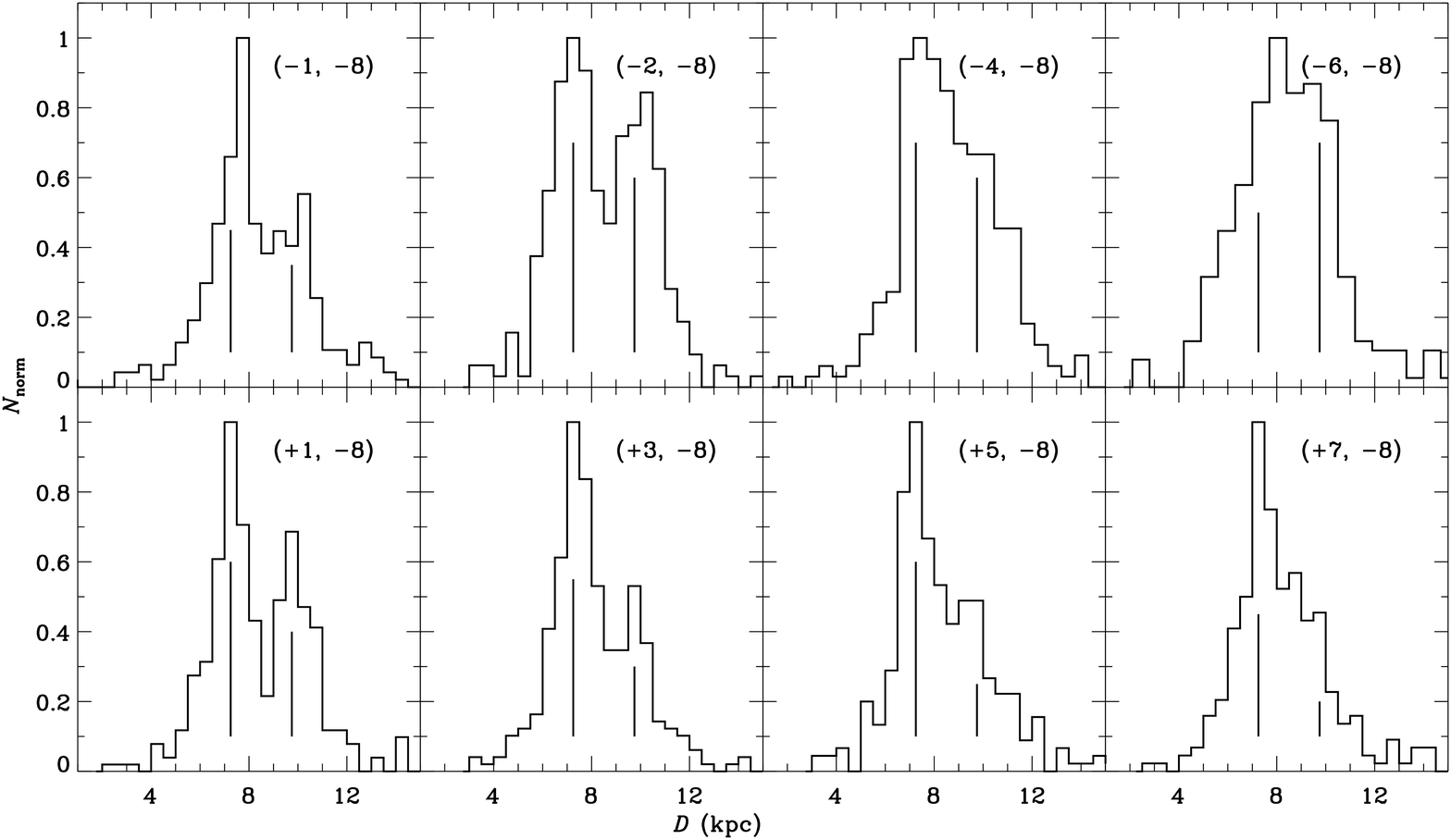}
\caption{Distance histograms of particles in fields with the same latitude ($b = -8^\degree$). The longitude ($l$) and latitude ($b$) of each box are shown in the upper right corner of each panel in the format $(l, b)$. The histograms have been normalized with the peak value as unity. This is to be compared to Figure~3 in MZ10. The positions of the double peaks in the field $(+1, -8)$ are marked with vertical lines (7.25 kpc and 9.75 kpc), which are also overplotted in other panels for comparison.}
\end{figure*}

\figurenum{3}
\begin{figure*}
\epsscale{0.9}
\plotone{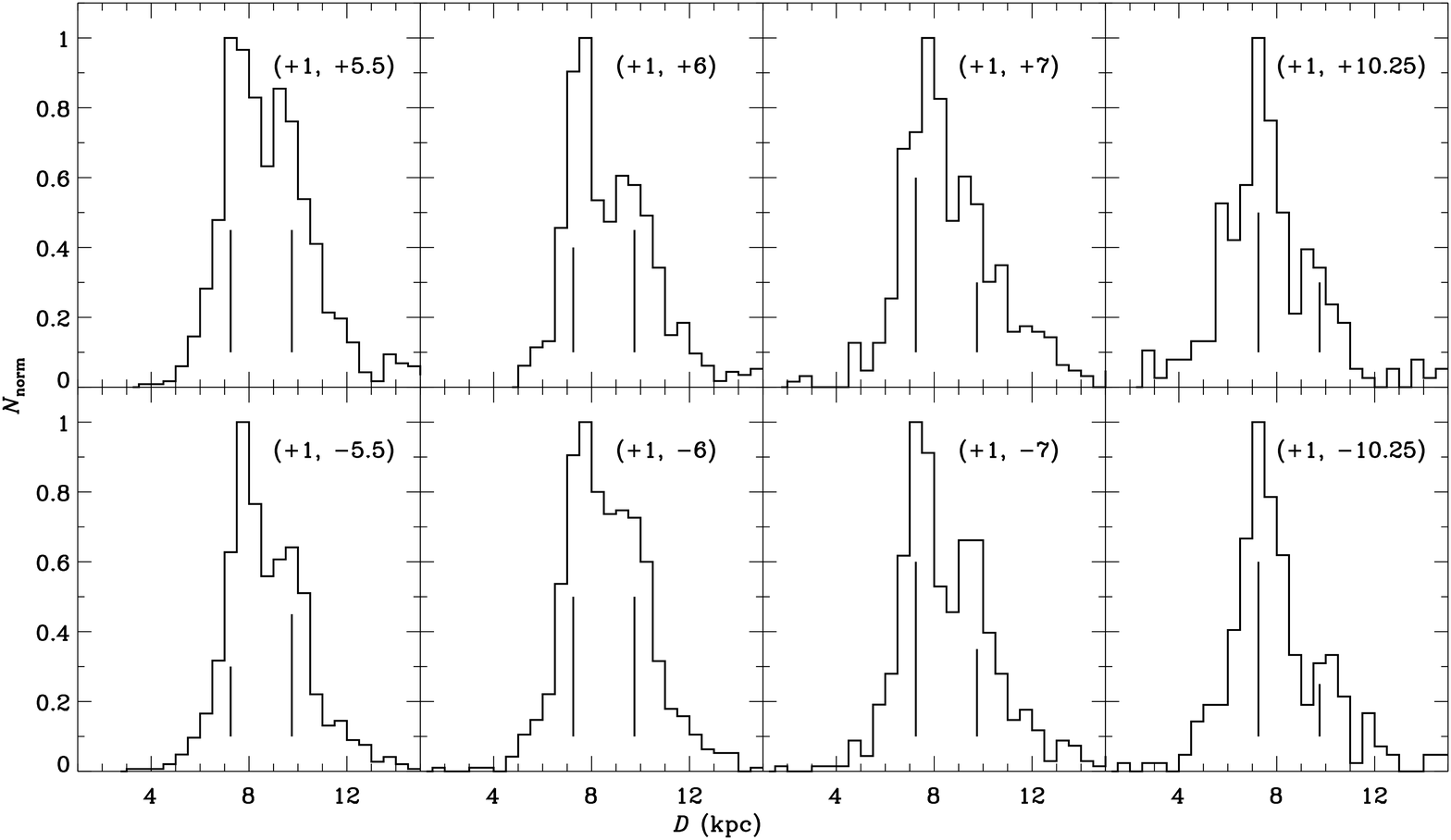}
\caption{Distance histograms of particles in fields with the same longitude ($l = +1^\degree$). The longitude ($l$) and latitude ($b$) of the box center are shown in the upper right corner of each panel in the format $(l, b)$. The histograms have been normalized with the peak value as unity. This is to be compared to Figure~7 in MZ10. The vertical lines in each panel mark the peak positions of the distance histogram in the field $(+1, -8)$}
\end{figure*}

\subsection{Comparison with observations}

It is of great interest to compare our model with the observational evidence of the X-shaped
structure in our Milky Way (MZ10; Nataf et al. 2010; S11). First, we need
to create the mock image in the solar perspective. The configuration is
exactly the same as Shen et al. (2010), where the Sun is 8.5 kpc
away from the GC ($R_{\odot} = 8.5\ \rm kpc$), and the Sun--GC line is $20^\degree$ away from the major axis of the bar. Note that the half length of the bar ($R_{\rm bar}$)  is about 4 kpc.

\subsubsection{Double peaks in distance histograms}

Figure~3 in MZ10 shows the luminosity functions of RC
stars in different fields at $b = -8^\degree$. A bimodal distribution shows up in almost all panels, which was also seen in Nataf et al. (2010). 
The relative amplitude of the bright to faint 
RC peaks changes dramatically with the longitude $l$; the faint RC dominate at negative
$l$ and the bright RC are more prominent at positive $l$. 
The two RCs at different $l$ actually have similar apparent 
magnitude in $K_0$  band, indicating that the distance to both RCs varies little with $l$. MZ10 found this result hard to understand with a naive straight bar.

Figure~2 shows the distance histograms of particles in different regions at $b = -8^\degree$.
To make a fair comparison, we study the fields identical to those in MZ10. Almost all panels in Figure~2 display a bimodal distribution. Note that the double peaks in the distance histograms actually correspond to those in the magnitude histograms (Figure~3 in MZ10); the brighter RC peak is closer to us, and the fainter RC peak is further away from us. The relative amplitude of the bright peak to the faint one decreases as the line of sight shifts from positive to negative $l$. 
However, at negative $l$, the front peak is still more prominent than the second peak, which is different from MZ10. This is mainly due to the fact that the front arm of the X-shaped structure in our model is wide in longitude at $b = -8^\degree$, covering from $+9^\degree$ to $-6^\degree$ in $l$. Moreover, in each field, particles at the front peak are closer to the Galactic plane than the second peak, therefore the space density at the front peak is higher.
The two vertical lines in each panel of Figure~2 mark the position of the double peaks in the field $(+1, -8)$, in the same fashion as MZ10.
We can find that the positions of the two peaks are roughly constant at different $l$ as in MZ10. 

Figure~3 shows the distance histograms of particles in fields at $l = +1^\degree$ with two vertical lines marking the same positions as in Figure~2 (to be compared to Figure~7 in MZ10). A bimodal distribution again shows up in all panels. For fields closer to the Galactic plane, the separation between the two peaks decreases, which is clear evidence for the X-shaped structure. The distance difference increases from $\sim$1.8 kpc at $b = \pm5.5^\degree$ to $\sim$2.5 kpc at $b = \pm10.25^\degree$. MZ10 also found that at lower latitude, the magnitude difference between the two peaks decreases, meaning the two RCs are getting close to each other. Figure~8 in MZ10 shows that the structure extends to about 3 kpc in the disk and 2 kpc in the vertical direction. The X-shaped structure in our model (Figure~1) has almost the same size.

So our model, which contains an X-shaped structure created naturally in the buckled bar, is in good agreement with the observational results of MZ10.

\subsubsection{Number density maps in latitude and longitude slices}

We compare the number density maps in latitude and longitude slices of our model to those obtained by S11. Figure~4 shows normalized stellar number densities for slices at different latitudes $b$. The projected GC is marked with a black cross at the center of each panel. Two overdensities show up in almost all panels, and the separation between them increases with $|b|$. The connection of the two overdensities tilts in a similar fashion as in S11; this is consistent with the fact that our bar angle ($20^\degree$) is the same as that found by S11. The two overdensities get closer as $|b|$ decreases towards the disk plane. Due to incompleteness of 2MASS at lower latitude ($|b| \leq 3.5^\degree$), S11 were unable to measure the number densities in those fields. We produce the stellar number density maps for slices at $|b| = 3^\degree$ and $2^\degree$ in our model. Our model predicts that the two overdensities are about 1 kpc apart at $|b| = 3^\degree$ (shown in Figure~4), and almost merge together for $|b| \leq 2^\degree$.

Another feature in Figure~3 of S11 is that the overdensity on the far side fades away faster than the closer overdensity. The same behavior is also seen in our Figure~4. Since the X-shaped structure looks symmetric in Figure~1, this behavior may be caused by the line-of-sight effect. At the same latitude, particles at large distances actually reside in diffuse ends of the X shape, so they fade faster than the closer overdensity region (S11). In Figure~4, the overdensity at larger distance becomes noisier for slices at $|b| \geq 6^\degree$ due to the limited number of particles in our model. We hope future higher resolution simulations will improve particle statistics on this.

Figure~5 shows the stellar number density maps of vertical slices at different $l$ toward the GC. A weak X shape is discernible for $|l| \leq 2^\degree$. This is qualitatively consistent with Figure~4 in S11. For slices at $|l| > 2^\degree$, the X-shaped structure becomes hard to identify as in S11.

\subsection{Origin and implications of the X-shaped structure}

The formation of the boxy bulge in our self-consistent model naturally produces a vertical X-shaped structure. The bar formation process enhances the radial streaming motion of stars along the bar, making the disk vulnerable to the buckling/firehose instability (Toomre 1966; also reviews by Sellwood \& Wilkinson 1993 and Sellwood 2010). As the instability gradually saturates, the thickened bar appears as a BPS bulge when viewed edge-on (Raha et al. 1991). 

The backbone orbits of a 3D buckled bar are the $x_1$ tree, i.e.,  the $x_1$ family plus a tree of 2D and 3D families bifurcating from it (Pfenniger \& Friedli 1991). 
The X-shaped structure is probably associated with orbits trapped around the 3D $x_1$ family, e.g., x1v1, x1v4 etc (e.g., Patsis et al. 2002; Athanassoula 2005). Note that the radial extent of our X-shaped structure is less than half of the bar length. The length of the bar is determined mainly by the $x_1$ family or 2D rectangular-like 4:1 resonance orbits, which have a larger radial extent (Patsis et al. 2003). On the other hand, the 3D backbone orbits of the X shape probably extend shorter in the radial direction (e.g., Patsis et al. 2002). 

This X-shaped structure does not have a straight-forward explanation in classical bulge formation scenarios (Bureau et al. 2006), but it is a natural consequence of the bar buckling mechanism as we show in this Letter. We can qualitatively reproduce the observational signatures of the X shape, such as double peaks in distance histograms (MZ10) and number density maps (S11). The existence of the X-shaped structure in our Milky Way may imply that the Galactic bulge is shaped mainly by internal disk dynamical instabilities instead of mergers.

De Propris et al. (2011) studied the radial velocity and abundances of bright and faint RCs at $(l, b) = (0^\degree, -8^\degree)$, and found no dynamical or chemical differences (also Uttenthaler et al. 2012). Proper motions of the two RCs are also similar (Vieira et al. 2007). These clues suggest that the two RCs indeed belong to the same coherent dynamical structure, which could be naturally made in the formation of the bar/boxy bulge.
 
In our $N$-body simulation, the buckling instability gradually saturates and the X-shaped structure becomes increasingly symmetric with time. We also studied the snapshot model at an earlier time of 2.4 Gyr. Unlike our canonical model at 4.8 Gyr, the X-shaped structure at 2.4 Gyr appears quite asymmetric about the disk plane. The observed symmetry (MZ10; S11) probably indicates that the X-shaped structure in the Galactic bulge has been in existence for at least a few billion years.

Although the X-shaped structure in our simple model is qualitatively similar to the observed one, it still cannot match all details of observations. For example, in the panels at negative longitudes in Figure~2, the second peak at larger distances is not as significant as the observed peak of the faint RC stars (Figure~3 of MZ10). Nevertheless, it is encouraging that our simple model matches observations in many aspects, and may help to guide future analyses. Further improvements on this model are clearly desired to completely understand the Galactic bulge structure, its dynamical and chemical histories.  

After this letter was published in the arXiv and submitted to ApJL, Ness et al. (2012) independently reported their study on the split red clump based on the ARGOS survey, and compared to an N-body boxy bulge model. Their main conlcusions are broadly consistent with ours.

\acknowledgements The research presented here is partially supported by the
National Natural Science Foundation of China under grant No. 11073037 to JS
and by 973 Program of China under grant No. 2009CB824800 to JS. We thank the anonymous referee for a constructive report. We are grateful to Michael Rich for encouraging this research, and we thank Jerry Sellwood, Andrea Kunder, Michael Rich, Danny Pan and Martin Smith for their helpful comments and discussions. JS wishes to acknowledge the hospitality of the Aspen Center for Physics, which is supported by the NSF grant No. 1066293, and where JS learned most details of the observed X-shaped structure.


\figurenum{4}
\begin{figure}
\plotone{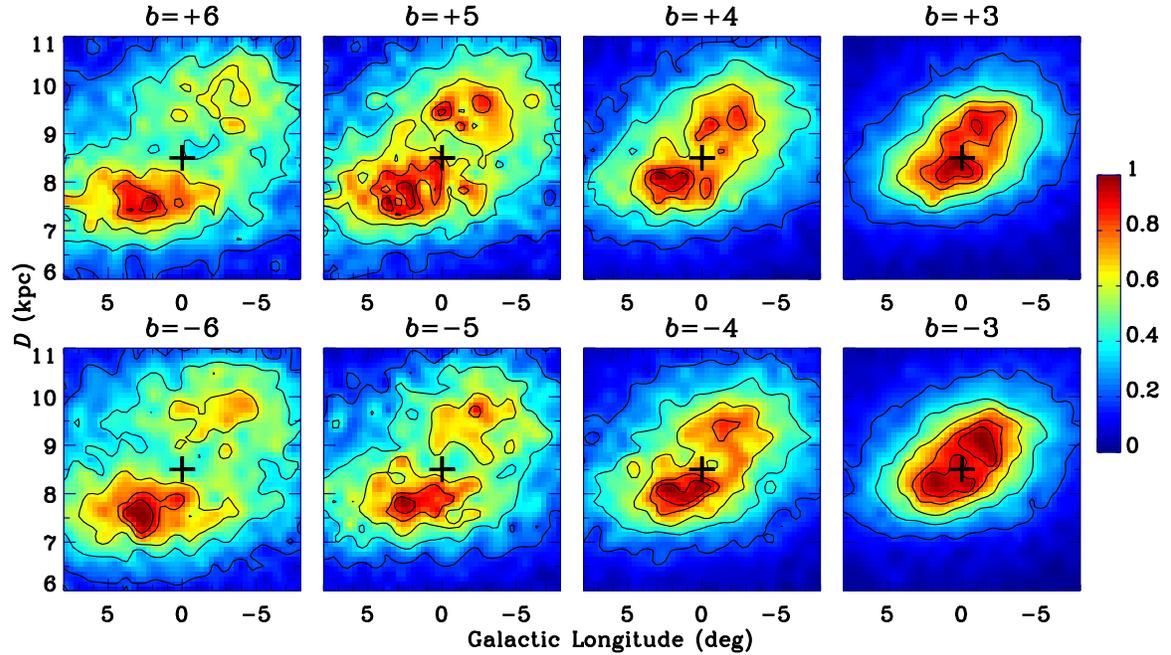}
\caption{Normalized stellar number density maps showing the structures toward the Galactic bulge. Each panel represents the density map of a particular latitude slice, which is labeled on top of the panel. The width of the slice is $1^\degree$. The black cross at the center of each panel marks the projected position of the GC. This is to be compared to Figure~3 in S11.}
\end{figure}

\figurenum{5}
\begin{figure}
\plotone{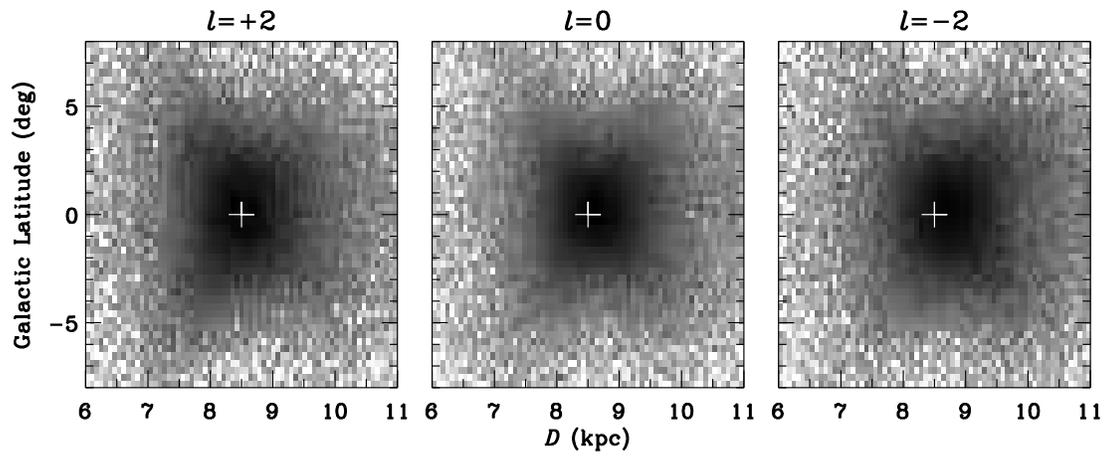}
\caption{Stellar number density maps toward the Galactic bulge at different longitude slices, which is labeled on top of each panel. Each slice is $1^\degree$ wide. The white cross at the center of each panel marks the projected position of the GC. This is to be compared to Figure~4 in S11.}
\end{figure}

\end{document}